\documentclass[a4paper,11pt,tightenlines,nofootinbib,american,aps,prd]{revtex4-1}

\usepackage[utf8]{inputenc}
\usepackage{babel}
\usepackage{hyperref}
\usepackage{amsmath}
\usepackage{cleveref}
\usepackage{graphics}
\usepackage{tabularx}

\newcommand{\calH}{\mathcal H}
\newcommand{\calA}{\mathcal A}
\newcommand{\calB}{\mathcal B}
\newcommand{\dd}{\mathrm{d}}
\newcommand{\keta}{\vec{k},\eta}

\newcommand{\qeta}{\vec{q},\eta}
\newcommand{\lcdm}{$\Lambda$CDM}

\usepackage{color}
\newenvironment{rc}{}{}

\begin{document}

\author{Adrian Vollmer}
\affiliation{Institut für Theoretische Physik, Ruprecht-Karls-Universität Heidelberg,
Philosophenweg 16, 69120 Heidelberg, Germany}

\author{Luca Amendola}
\affiliation{Institut für Theoretische Physik, Ruprecht-Karls-Universität Heidelberg,
Philosophenweg 16, 69120 Heidelberg, Germany}

\author{Riccardo Catena}
\affiliation{Institut f\"ur Theoretische Physik, Georg-August-Universit\"at G\"ottingen,
Friedrich-Hund-Platz 1, 37077 G\"ottingen, Germany}

\title{Efficient implementation of the Time Renormalization Group}

\begin{abstract} {\rc The Time Renormalization Group (TRG) is an effective
method for accurate calculations of the matter power spectrum at the scale
of the first baryonic acoustic oscillations.} By using a particular variable
transformation in the TRG formalism, we can reduce the 2D integral in the
source term of the equations of motion for the power spectrum into a series
of 1D integrals. The shape of the integrand allows us to pre-compute only
thirteen antiderivatives numerically, which can then be reused when
evaluating the outer integral.  While this introduces a few challenges to
keep numerical noise under control, we find that the computation time for
nonlinear corrections to the matter power spectrum decreases by a factor of
50. This opens up the {possibility to use of TRG for mass production as in 
Markov Chain Monte Carlo methods}. A Fortran code demonstrating this new
algorithm has been made publicly available.  \end{abstract}

\maketitle

\section{Introduction}

Future observations of the large scale structure of the Universe are
expected to place constraints on cosmological parameters {\rc of fundamental
importance}. The ESA spacecraft Euclid \cite{Laureijs2011,Amendola2013b},
for example, is going to measure the ellipticity and the redshift of
billions of galaxies, providing us with a measurement of the weak lensing
convergence power spectrum, which is sensitive to {\rc currently unknown}
dark energy parameters \cite{Huterer2002,Amara2007}. To make forecasts on
the constraints {\rc expected for} these parameters, {\rc and to interpret
real data in the future}, we typically need to sample the convergence power
spectrum, and thus the matter power spectrum, at many different points in
the parameter space of a given cosmological model. In particular, we need
the matter power spectrum at scales and redshifts where nonlinear
corrections {\rc to the fluid equations} become important.

{\rc Methods to calculate the matter power spectrum in the nonlinear regime
of cosmic perturbations are hence of prime importance.} {\rc The mildly
nonlinear cosmic scales are of particular interest in the study of models
beyond \lcdm, as they correspond to the first baryonic acoustic oscillations
(BAO) probed by Euclid.} {\rc There are complementary strategies to compute
the matter power spectrum down to redshift zero at these relatively small
scales.} These include using a phenomenological fitting function as in
\cite{Eisenstein1999} and performing $N$-body simulations.  However, those
methods have significant drawbacks. $N$-body simulations require a large
amount of CPUs and memory and are still too time consuming to be practical
{\rc in statistical parameter inference}, {where typically hundreds of thousands combinations of
parameter choices are needed}.  Fitting functions are not based
on first principles and are quite limited in regard to applicable
cosmological models.

Another way to obtain the matter power spectrum {\rc at the BAO scale} is to
solve the fluid equations perturbatively. {\rc Significant progress has
been made in this context in the last few years. As a result, the
limitations of Standard Perturbation Theory (SPT) (see~\cite{Bernardeau2002}
for an extensive review) have become increasingly more
clear~\cite{Crocce:2005xy}. In SPT, perturbative corrections to the matter
power spectrum of different order in the density contrast are of comparable
size on nonlinear scales and can have opposite signs, with significant
cancellations among different terms in the perturbative
expansion~\cite{Crocce:2005xy}. This ``instability'' of the theory implies
poor convergence properties for the perturbative series: many terms in the
perturbative expansion of the matter power spectrum have to be considered in
order to obtain an accurate prediction. 

In Renormalized Perturbation Theory
(RPT)~\cite{Crocce:2005xy,Crocce:2005xz}, the SPT perturbative expansion of
the power spectrum is reorganized by resumming an infinite class of terms in
the perturbative series through a diagrammatic approach. As a result, in RPT
the perturbative series is composed of positive terms only, and successive
perturbative corrections to the matter power spectrum dominate at
increasingly smaller scales. Thereby, in RPT the perturbative series
converges much more efficiently than in SPT, and just few terms are required
in order to accurately predict the matter power spectrum on mildly nonlinear
scales~\cite{Crocce:2007dt}. 

The ability of RPT in fitting results of $N$-body simulations has motivated
the exploration of similar resummation schemes. The Time Renormalization
Group (TRG) approach~\cite{Pietroni2008}, for instance, is a resummation
scheme widely explored in the past few years. It resums all perturbative
corrections to the matter power spectrum in which the so-called
``interaction vertex'', i.e. a matrix encoding the coupling of different
physical scales induced by nonlinear effects, is kept at its tree level
form. In contrast to other methods, the TRG approach has the advantage of
evaluating the time dependence of the matter power spectrum {\it exactly}
through a set of differential equations in the time variable. In addition,
it can be easily applied to a broad class of cosmological
models~\cite{Lesgourgues:2009am,Saracco:2009df}. Alternative resummation
schemes rely on renormalization group equations~\cite{Matarrese:2007wc},
effective field theory
methods~\cite{Baumann:2010tm,Carrasco:2012cv,Piazza:2013coa},
multi-propagator expansions~\cite{Bernardeau:2008fa,Bernardeau:2011dp}, the
eikonal approximation~\cite{Bernardeau:2011vy}, and Lagrangian perturbation
theory~\cite{Matsubara:2007wj,Bernardeau:2008ss}. Recently, the possibility
of combining perturbation theory with $N$-body simulations has also been
explored~\cite{Pietroni:2011iz,Manzotti:2014loa}.}

While existing {\rc numerical} implementations of the TRG are already many
orders of magnitudes faster than $N$-body simulations, it is still not quite
there for practical applications. Their runtime is of the order of one hour
on a typical desktop machine, {\rc which makes extensive explorations of
multidimensional parameter spaces unfeasible.} As we will show in this
article, we managed to exploit a feature of the integrand in the source term
of the TRG equations that enables us to reduce the runtime to less than a
minute or even a few seconds, depending on the CPU power of the computer.
This opens up the possibility of using {\rc higher order} perturbation
theory, {\rc and the TRG}, for applications like weak lensing Fisher matrix
analysis, {\rc and Markov Chan Monte Carlo (MCMC) explorations of} models
beyond \lcdm. {\rc Efficient numerical implementations of other resummation
schemes can be found in~\cite{Crocce2012,Taruya:2012ut}.}

The paper is structured as follows. In \cref{sec:trg}, we briefly review the
TRG framework as described in \cite{Pietroni2008} and outline the derivation
of the TRG equations. We elaborate on the problems involved with
implementing a numerical algorithm that solves the TRG equations and how to
solve them in \cref{sec:numerics}.  \Cref{sec:analysis} contains a
comparison with existing implementations in Copter \cite{Carlson2009} and an
older version of \textsc{Class} \cite{Audren2011} as well as $N$-body simulations.
Finally, we present our conclusions in \cref{sec:conclusion}.

The Fortran code implementing this method, which we call \textsc{trgfast},
can be downloaded at \url{https://gitub.com/User0815/trgfast}.

\section{Review of TRG}
\label{sec:trg}

The goal is to solve the fluid equations,
\begin{align}
\frac{\partial \delta_\mathrm{m}}{\partial \tau} + \vec \nabla \cdot ( (
1+\delta_\mathrm{m}) \vec v)& = 0 \,, \label{eq:cont}\\
\frac{\partial \vec v}{\partial \tau} + \calH(\vec v + [\calA \vec
v]) + (\vec v\cdot \vec\nabla)\vec v& = -\vec\nabla\phi\,,\label{eq:euler}\\
 \frac{3}{2} \calH^2\Omega_\mathrm{m}(\delta_\mathrm{m} +
[\calB \delta_\mathrm{m}]) &= \nabla^2\phi\,, \label{eq:poisson}
\end{align}
which consist of the continuity equation, the Euler equation, and the
Poisson equation, respectively. Here, $\delta_\mathrm{m}$ is the matter
density contrast, $\vec v$ the comoving velocity field, $\Omega_\mathrm{m}$
the time dependent average matter density in units of the critical density,
and $\phi$ the Newtonian gravitational potential. The brackets denote
convolution.  Note the presence of two extra terms that allow for more general
cosmological models, represented by
the functions $\calA(\vec k, \tau)$ and $\calB(\vec k, \tau)$.  These are
always non-zero when the geodesics of particles are modified, for example in
Brans-Dicke  cosmologies \cite{Pietroni2008}, massive neutrinos
\cite{Lesgourgues2006,Anselmi2011}, etc. As usual in cosmological
perturbation theory, we define the velocity divergence $\theta = \vec \nabla
\cdot \vec v$ and neglect the vorticity $\vec \nabla\times \vec v$. 

Writing the fluid equations in Fourier space yields
\begin{align}
\label{eq:fluidfourier1}
\frac{\partial \delta_\mathrm{m}(\vec k, \tau)}{\partial \tau} + \theta(\vec k, \tau)
+ \int\dd^3\vec q \dd^3\vec p~\delta_\mathrm{D}(\vec k - \vec q - \vec p)
\alpha(\vec q, \vec p) \theta(\vec q,\tau) \delta_\mathrm{m}(\vec p,\tau) &=
0\,,\\
\frac{\partial\theta(\vec k,\tau)}{\partial \tau}+\calH(1+\calA(\vec k,
\tau))\theta(\vec k, \tau) + \frac32 \calH^2(1+\calB(\vec k, \tau))
\Omega_\mathrm{m}(\tau) \delta_\mathrm{m}(\vec k, \tau) & \nonumber \\
+ \int \dd^3\vec q  \dd^3\vec p~\delta_\mathrm{D}(\vec k -\vec q-\vec p)\beta(\vec
q,\vec p)\theta(\vec q,\tau) \theta(\vec p, \tau)  & = 0\,,
\label{eq:fluidfourier}
\end{align}
where the nonlinearity is encoded in the two mode coupling terms,
\begin{equation}
\label{eq:modecoup}
\alpha(\vec k_2, \vec k_1) \equiv \frac{(\vec k_1 + \vec k_2)\cdot \vec k_2}{k_2^2},
\qquad \beta(\vec k_2, \vec k_1) \equiv\frac{(\vec k_1 + \vec k_2)^2\vec k_1 \cdot \vec
k_2}{2k_1^2k_2^2}\,. 
\end{equation}
These equations are often written in a more compact way by defining the
doublet
\begin{equation} 
\left(
\begin{array}{c}
\varphi_1(\vec k,\eta) \\
\varphi_2(\vec k,\eta)
\end{array}
\right) \equiv e^{-\eta} \left(
\begin{array}{c}
\delta_\mathrm{m}(\vec k, \eta)\\
-\theta(\vec k , \eta)/\calH
\end{array}
\right)\,, 
\label{eq:doubletdef} 
\end{equation}
where $\eta$ is the e-folding time, defined as
\begin{equation}
\eta \equiv \log \frac{a}{a_\mathrm{ini}}\,.
\label{eq:etadef}
\end{equation}
The initial scale factor is arbitrary in principle, but should be chosen to
be well inside the linear regime. Typical are values that correspond to a
redshift between $z=35$ and $z=100$. 
We also define a matrix that encapsulates the background evolution as
\begin{equation}
\Omega(\vec k, \eta) = \left(
\begin{array}{cc}
1 & -1 \\
-\frac{3}{2}\Omega_\mathrm{m}(\eta) (1+\calB(\vec k,\eta)) & 2+\frac{\calH'}{\calH} +
\calA(\vec k,\eta)
\end{array}
\right )\,.  \label{eq:bgOmega}
\end{equation}
These definitions allow us to write 
\cref{eq:fluidfourier,eq:fluidfourier1} in the form
\begin{align}
\partial_\eta \varphi_a(\vec k, \eta) =& - \Omega_{ab}(\vec
k,\eta)\varphi_b(\vec k, \eta) \nonumber\\
&+e^\eta\int \dd^3\vec q\dd^3\vec p~\gamma_{abc} 
(\vec k,-\vec p,-\vec q)\varphi_b(\vec p, \eta) \varphi_c(\vec q,\eta)\,,
\label{eq:dphia}
\end{align}
where the non-vanishing components of the vertex function $\gamma_{abc}$ read
\begin{align}
\label{eq:gammavert}
 \gamma_{121}(\vec k, \vec p, \vec q) &= \frac{1}{2}\delta_\mathrm{D}(\vec k+\vec p
+ \vec q)\alpha(\vec p, \vec q)\,,\\
 \gamma_{222}(\vec k, \vec p, \vec q) &= \delta_\mathrm{D}(\vec k+\vec p
+ \vec q)\beta(\vec p, \vec q)\,,\\
 \gamma_{121}(\vec k, \vec p, \vec q) &= \gamma_{112}(\vec k, \vec q, \vec
 p)\,.
\end{align}
Now we can write down an infinite series of equations that give us the
evolution of the power spectrum of $\varphi_a$. They can be obtained by
applying the product rule on the left-hand-side and plugging in
\cref{eq:dphia}. When suppressing the momentum dependence (integration over
$\vec q$ and $\vec p$ is understood), we get
\begin{align}
\partial_\eta\,\langle \varphi_a \varphi_b\rangle =& -\Omega_{ac} 
\langle \varphi_c \varphi_b\rangle   - 
\Omega_{bc} 
\langle \varphi_a \varphi_c\rangle 
\nonumber\\
&+e^\eta \gamma_{acd}\langle \varphi_c\varphi_d \varphi_b\rangle +e^\eta
\gamma_{bcd}\langle \varphi_a\varphi_c \varphi_d\rangle\,,\nonumber\\
\partial_\eta\,\langle \varphi_a \varphi_b  \varphi_c \rangle  =&
-\Omega_{ad} 
\langle \varphi_d \varphi_b\varphi_c\rangle  -\Omega_{bd} 
\langle \varphi_a \varphi_d\varphi_c\rangle  -\Omega_{cd} 
\langle \varphi_a \varphi_b\varphi_d\rangle \nonumber\\
&+e^\eta \gamma_{ade}\langle \varphi_d\varphi_e \varphi_b\varphi_c\rangle  
+e^\eta \gamma_{bde}\langle \varphi_a\varphi_d \varphi_e\varphi_c\rangle
\nonumber\\
&+e^\eta \gamma_{cde}\langle \varphi_a\varphi_b \varphi_d\varphi_e\rangle
\,, &\nonumber\\
\partial_\eta\,\langle \varphi_a \varphi_b  \varphi_c  \varphi_d \rangle
=&~\cdots\nonumber\\
\vdots \label{eq:tower}
\end{align}
The definition of the power spectrum $P_{ab}(\vec k, \eta)$, the bispectrum
$B_{abc}(\vec k, \vec q, \vec p, \eta)$ and
the connected part of the trispectrum $Q_{abcd}(\vec k, \vec q, \vec p, \vec
r, \eta)$, respectively, is
\begin{align}
\langle \varphi_a({\bf k},\,\eta) \varphi_b({\bf q},\,\eta)&\rangle
\equiv \delta_\mathrm{D}({\bf k + q}) P_{ab}({\bf k}\,,\eta)\,,\nonumber\\
\langle \varphi_a({\bf k},\,\eta) \varphi_b({\bf
q},\,\eta)&\varphi_c({\bf p},\,\eta)\rangle \equiv \delta_\mathrm{D}({\bf k +
q+p})
B_{abc}({\bf k},\,{\bf q},\,{\bf p};\,\eta)\,,\nonumber\\
\langle \varphi_a({\bf k},\,\eta) \varphi_b({\bf
q},\,\eta)&\varphi_c({\bf p},\,\eta)\varphi_d({\bf r},\,\eta)\rangle
\equiv\nonumber\\
&   \left[\delta_\mathrm{D}({\bf k + q })\,
\delta_\mathrm{D}({\bf p + r }) P_{ab}({\bf k}\,,\eta)P_{cd}({\bf
p}\,,\eta)\right.\nonumber\\
& +\delta_\mathrm{D}({\bf k + p}) \,\delta_\mathrm{D}({\bf q + r
}) P_{ac}({\bf k}\,,\eta)P_{bd}({\bf q}\,,\eta)\nonumber\\
& +\delta_\mathrm{D}({\bf k + r})\, \delta_\mathrm{D}({\bf q + p
}) P_{ad}({\bf k}\,,\eta)P_{bc}({\bf q}\,,\eta)\nonumber\\
& \left. +\,\delta_\mathrm{D}({\bf k + p+q+ r})
\,Q_{abcd}({\bf k}\,,{\bf q}\,,{\bf p}\,,{\bf r}\,,\eta)\right]\,.
\end{align}
The four-point function is partially written in terms of the two-point
functions by using the Wick theorem, but since we allow for a non-zero
bispectrum, the field $\varphi_a$ does not need to be Gaussian~\cite{Bartolo:2009rb}. The
approximation we make in TRG is to set $Q_{abcd} = 0$. This closes the
system of equations for the evolution of the power spectrum and from
\cref{eq:dphia} we get 
\begin{align} 
\partial_{\eta}P_{ab}(\keta)=&-\Omega_{ac}(\keta)P_{cb}(\keta)
-\Omega_{bc}(\keta)P_{ac}(\keta)\nonumber\\
&+e^{\eta}\int
\dd ^3\vec{q}\left[\gamma_{acd}(\vec{k},-\vec{q},\vec{q}-
\vec{k})B_{bcd}(\vec{k},-\vec{q},
\vec{q}-\vec{k};\eta)\right.\nonumber\\
&\hspace{1.2cm}\left.+B_{acd}(\vec{k},-\vec{q},
\vec{q}-\vec{k};\eta)\gamma_{bcd}(\vec{k},-\vec{q},
\vec{q}-\vec{k},\eta)\right]\,,\label{eq:dpab}
\end{align}
\begin{align}
\partial_{\eta}B_{abc}(\vec{k},-\vec{q}, \vec{q}-\vec{k};\eta) = &
-\Omega_{ad}(\keta)B_{dbc}(\vec{k},-\vec{q},
\vec{q}-\vec{k};\eta)\nonumber\\
&-\Omega_{bd}(-\vec{q},\eta)B_{adc}(\vec{k},-\vec{q},
\vec{q}-\vec{k};\eta)\nonumber\\
&-\Omega_{cd}(\vec{q}-\vec{k},\eta)B_{abd}(\vec{k},-\vec{q},
\vec{q}-\vec{k};\eta)\nonumber\\
&+2e^{\eta}\left[\gamma_{ade}(\vec{k},-\vec{q},
\vec{q}-\vec{k})P_{db}(\vec{q},\eta)P_{ec}(\vec{k}-\vec{q},\eta)\right.\nonumber\\
& +\gamma_{bde}(-\vec{q},
\vec{q}-\vec{k},\vec{k})P_{dc}(\vec{k}-\vec{q},\eta)P_{ea}(\vec{k},\eta)\nonumber\\
& +\left.\gamma_{cde}(\vec{q}-\vec{k},\vec{k},-
\vec{q})P_{da}(\keta)P_{eb}(\qeta)\right]\,.\label{eq:dbabc}
\end{align}
However, the equations in this form are unsuitable for numerical
computation, mainly because the bispectrum depends on three momenta, which
would require a large 3D array to store in memory. 
Rewriting \cref{eq:dpab,eq:dbabc} to reflect the isotropy of the Universe,
we get
\begin{align}
\partial_{\eta}P_{ab}(k,\eta)=&-\Omega_{ac}(k,\eta)P_{cb}(k,\eta)-
\Omega_{bc}(k,\eta)P_{ac}(k,\eta)\nonumber\\
&+e^{\eta}\frac{4\pi}{k}\int_{k/2}^{\infty}\hspace{-0.3cm}q~\dd q
\int_{|q-k|}^q\hspace{-0.3cm}p~\dd p\left[\tilde\gamma_{acd}(k,q,p)
\tilde B_{bcd}(k,q,p;\eta)\right.\nonumber\\
&\hspace{2cm}\left.+\tilde B_{acd}(k,q,p;\eta)\tilde\gamma_{bcd}(k,q,p)
\right]\,,\label{eq:dpab2} 
\end{align}
where we defined
\begin{equation}
\tilde \gamma_{abc}(k,q,p) \equiv \gamma_{abc}(\vec{k},\vec{q},\vec{p})|_{\vec{p}=
-(\vec{k}+\vec{q})}
\end{equation}
and analogously for $\tilde B_{abc}(k,q,p)$. The vertex functions are then
equivalent to 
\begin{align}
\tilde\gamma_{121}(k,q,p) = &\frac{q^2+k^2-p^2}{4q^2}\,,\\
\tilde\gamma_{222}(k,q,p)= &\frac{k^2}{4p^2q^2}(k^2-p^2-q^2)\,.
\end{align}
Since here we are not actually interested in the bispectrum, and two out of the
three momenta on which the bispectrum depends on are integrated out in
\cref{eq:dpab2}, we can do the same integration on both sides of
\cref{eq:dbabc}.
Then it becomes convenient to define a quantity that directly depends on the
bispectrum,
\begin{align}
I_{acd,bef}(k)&\equiv\int_{k/2}^{\infty}\hspace{-0.3cm}\dd q~q
\int_{|q-k|}^q\hspace{-0.3cm}\dd p~p\frac12\left[
\tilde\gamma_{acd}(k,q,p)\tilde B_{bef}(k,q,p)+(q\leftrightarrow
p)\right]\,,
\label{eq:defI}
\end{align}
such that \cref{eq:dpab2,eq:dbabc} can be written as
\begin{align}
\partial_{\eta}P_{ab}(k)=&-\Omega_{ac}P_{cb}(k)-\Omega_{bc} 
P_{ac}(k)+e^{\eta}\frac{4\pi}{k}
[I_{acd,bcd}(k)+I_{bcd,acd}(k)]\label{eq:dpabI}\\
\partial_{\eta}I_{acd,bef}(k)=&-\Omega_{bg}I_{acd,gef}(k)
-\Omega_{eg}I_{acd,bgf}(k)\nonumber \\ &-\Omega_{fg}I_{acd,beg}(k) +
2e^{\eta}A_{acd,bef}(k)\,.\label{eq:dII}
\end{align}
Here, we defined another quantity: the source term 
\begin{align} 
A_{acd,bef}(k)\equiv \int_{k/2}^{\infty}\hspace{-0.3cm}\mathrm
dq~q\int_{|q-k|}^{q}\hspace{-0.3cm}\mathrm dp~p\frac12
\left\{\tilde\gamma_{acd}(k,q,p)\left[\tilde\gamma_{bgh}(k,q,p)P_{ge}(q)P_{hf}(p)
+\right.\right.\nonumber\\[.3cm]
\left.\left.\tilde\gamma_{egh}(q,p,k)P_{gf}(p)P_{hb}(k)+
\tilde\gamma_{fgh}(p,k,q)P_{gb}(k)P_{he}(q)\right]
+ (q\leftrightarrow p)\right\}\,.\label{eq:defA} 
\end{align}
At this point we also define the integrand of the outer integral via
\begin{equation}
\label{eq:Ak}
A_{acd,bef}(k) \equiv \int_{k/2}^{\infty}\hspace{-0.3cm}\dd
q~K_{acd,bef}(k,q)\,,
\end{equation}
which will become convenient in the next section. In the original paper
\cite{Pietroni2008}, it is recommended to perform a transformation of
variables to make the integral limits of the inner integral independent of
the integration variable $q$, but as we will see it is beneficial to keep
the shape of $A_{acd,bef}$ as in \cref{eq:defA}.

After considering all index symmetries of $A_{acd,bef}$, it can be deduced
that there are only 14 non-vanishing, independent components. Together with
the three independent components of $P_{ab}$, this means that
there are 17 independent equations in total in the system
\cref{eq:dpabI,eq:dII}.

\section{Implementation}
\label{sec:numerics}

The integral in \cref{eq:defA} clearly provides the bottleneck when solving
\cref{eq:dpabI,eq:dII}, as it represents the nonlinear part in an otherwise
regular system of ordinary differential equations. Performing the 2D integral
is quite costly in terms of CPU time. However, a close inspection of the
integral in \cref{eq:defA} (after expanding it with the help of Mathematica)
shows that the integrand can be written in terms of only 13 expressions,
which we will call \emph{moments}, of the shape 
\begin{equation}
p^m P_{ab}(p)\,,
\end{equation}
with a pre-factor that depends only on $q$, $k$ and $\eta$. Here, $m$ is an
integer out of the set ${-3,-1,1,3,5}$. The third term in \cref{eq:defA} can
even be integrated analytically, as it does not depend on $P_{ab}(p)$ and
the vertex function $\tilde\gamma_{fgh}(p,k,q)$ is a rational function of
$p$.

This allows us to separately perform the $p$-integration first and then do
the $q$-integration, reducing the 2D integral into a relatively small number
of 1D integrals. However, not only do we have to compute the antiderivatives
of the 13 moments, we also need to compute the difference of these
antiderivatives at $q$ and $|q-k|$. Differences between two terms are
notoriously difficult to compute when the terms are almost equal. This
problem is amplified by the fact that all non-analytic quantities (such as
the power spectrum) are represented by cubic splines when numerically
evaluating the expressions considered here. Since there are terms containing
up to seventh powers of $q$, even small finite differences between terms
that should cancel exactly become very significant.  To avoid the
amplification of numerical artifacts, we use the following techniques.

\subsection{Spline integration}

When performing numerical calculations, any function that does not have an
analytic expression (such as the power spectrum) is typically given by an
ordered set of tuples, which we will call the sampling points. If such a
function is to be evaluated between two sampling points, we can only
approximate the true value of the function at that value by interpolating
it.  In this application, we choose natural cubic splines. Thus, the moments
are the product of a power of $p$ and a function given by cubic splines. 
We found that we get the least amount of numerical noise if we interpolate
the power spectrum in log-log-space, i.e. instead of $P(k)$ we actually
interpolate $\log P(\exp(x))$ with $x=\log k$. This means we need to take
the logarithm of the wavenumber when evaluating the interpolated power
spectrum and apply the exponential function to the result. While this leads
to many calls to the \texttt{exp} and \texttt{log} functions and to an
increase in CPU time, it is necessary to make the code more stable. However,
when integrating the moments, we found it is best to interpolate
$P(\exp(x))$, as this enables us to perform the integration analytically in
each spline interval.

The integration constant for each interval needs to be chosen such that the
result is continuous at the sampling points. The integration constant for
the first interval is arbitrary in principle, but in a numerical context
this choice can be crucial.
In our case, the integration constant can be chosen such that the integrated
moment goes to zero either for small $k$ or for large $k$, depending on the
value of $m$. By doing so, we avoid nonzero asymptotes, which tend to lead
to catastrophic cancellation when subtracting two values of the
antiderivative from each other. ``Catastrophic cancellation'' refers to the
loss of precision when numerically computing differences of nearby values.
By choosing the integration constant such that the antiderivative of a
moment is zero at the first sampling point for $m \leq 1$ and zero at the
last sampling point otherwise, the numerical noise in the time
evolution of the power spectrum reduces substantially.

\subsection{Custom sampling}

After computing the antiderivatives numerically, all that is left to do is
to perform the $q$-integration in \cref{eq:defA}. Since the general shape of
the integrands is always the same, it is possible to use the simple
trapezoidal rule instead of an adaptive integration algorithm. The
disadvantage is that we have no error estimate on the result of the
integration, but at the same time it is faster to apply the trapezoidal
rule. To ensure we achieve sufficient accuracy, we verify the final result
by comparing it to the power spectrum obtained from both \textsc{Class} and $N$-body
simulations (see \cref{sec:analysis}).

\begin{figure}[tbp]
\begin{center}
\includegraphics{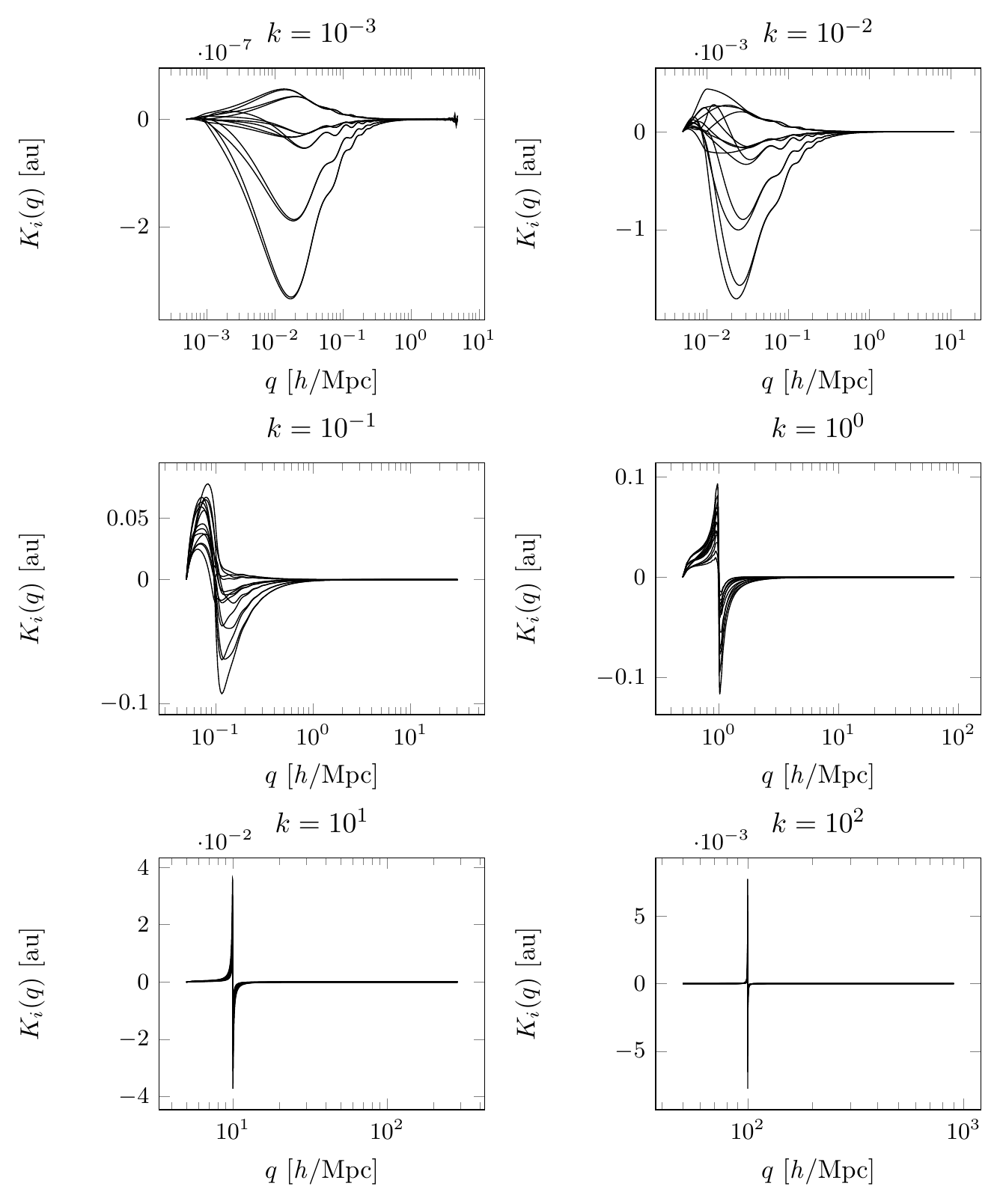}
\end{center}
\caption{All 14 integrands (in arbitrary units) after carrying out the
$p$-integration at different values for $k$. As $k$ increases, the
integrands approach a function with a vertical asymptote and a singularity
at $k$.}
\label{fig:Kdq}
\end{figure}

When plotting the
integrands for different $k$ (\cref{fig:Kdq}), we can observe that they are
all well-behaved for low $k$, but start to look increasingly similar to the
function $f(q)=1/(k-q)$ as $k$ enters the nonlinear regime.  Note, however,
that the function is continuous at all times and does have a zero near
$q=k$, it just changes its sign very rapidly. Since a lot of cancellation
happens in this region, it is important that we make sure to get an accurate
estimate of the integral there.

We use the knowledge we have about the shape of the integrand to restrict the sampling
scheme to two classes depending on some threshold $k_\mathrm{th}$: For
$k<k_\mathrm{th}$ we use logarithmically equidistant points between $k/2$
and $k_\mathrm{max}$ (the maximum wavenumber at which the power spectrum is
defined), and for $k>k_\mathrm{th}$ we first translate the
function to the left by transforming $q\to q-k$. Then we use logarithmically
equidistant points from $0$ to $-k/2$ and from $0$ to $k_\mathrm{max}-k$
before undoing the transform. We refer to these
sampling schemes as ``asymmetric scheme'' and ``symmetric scheme''
respectively. This way the sampling density is much higher where the
curvature of the integrand is large, yielding a good approximation of the
real value of the integral.  Of course, a suitable value for the threshold
$k_\mathrm{th}$ needs to be determined empirically.

\subsection{Cutoff method}

Since the upper limit of the integral in \cref{eq:Ak} is infinity, we would
have to evaluate the antiderivatives outside of their domain. There are a
few options on how to account for this: Setting the power spectrum to zero,
keeping it constant after the last defined sampling point, or extrapolating
it smoothly via a power law or exponential law. We found that the type of
extrapolation has no major impact on the final result.

Another issue are numerical instabilities, which cause the UV end of the
power spectrum to oscillate. As soon as these oscillations are introduced,
they quickly amplify and cause one of the three power spectra to take on
negative values. 

Even when these oscillations are absent, as the power spectrum evolves, the
$P_{22}$ component takes on very small values, approaching zero as the time
evolution progresses. These numerical artifacts  are the manifestation of a
limitation of the fluid equation themselves, which neglect the vorticity of
the velocity field and the velocity dispersion \cite{Pueblas2009,Valageas2010}.

To work around these issues, we decrease the maximum wavenumber
$k_\mathrm{max}$ at which the power spectrum is defined as the time $\eta$
increases.  The authors of \cite{Audren2011} did it for \textsc{Class} by
employing a technique they call ``double escape'', which essentially
consists of throwing out the last four points in each time step. In our
implementation, however, it is handled differently, since the number and
positions of the sampling points are given by the user.  After all, if the
algorithm works in a stable way for one set of sampling points, it should
yield the same result if the density of the sampling points is doubled, for
instance, which would not be the case if we would only throw out the last
two sampling points in each time step.  We will now describe how we
determine $k_\mathrm{max}$ at each time step.

First we tackle the numerical noise.  Since we want to avoid uneven behavior
such as oscillations and kinks in the power spectrum, we discard sampling
points at the UV end one by one until the curvature between the last three
sampling points of all three components varies only by a limited amount.
This happens at each time step. Conveniently, the coefficient of the
quadratic term in the splines gives a good
measure of the curvature.  In particular, we demand that the standard
deviation of the coefficients of the quadratic term in the last two spline
intervals of all three components of the power spectrum does not exceed a
certain limit $\sigma_c$.

The other issue of one of the power spectrum components becoming negative
can be avoided by imposing that the magnitude of the slope at the UV end of
$P_{22}(k)$ stays below another limit, which we will call $s_{22}$.  The
optimal values for $\sigma_c$ and $s_{22}$ have been determined in
empirically.

During the time evolution, we then simply set 
\begin{equation}
\partial_\eta P_{ab}(k) = 0
\end{equation}
and 
\begin{equation}
\partial_\eta I_{acd,bef}(k) = 0
\end{equation}
if $k>k_\mathrm{max}$, which takes these scales effectively out of the
equation. This cutoff method appears to lead to a stable time evolution
without discarding too much information. At the last time step, when $\eta$
corresponds to $z=0$, we typically have $k_\mathrm{max} \approx 0.8
h/\mathrm{Mpc}$.

\section{Analysis}
\label{sec:analysis}

\subsection{Class and Copter}

We ran the code for a flat \lcdm\ Universe with $\Omega_m=0.3175$,
$\Omega_b=0.0490$, $h=0.6711$, $\sigma_8=0.8344$, $n_s=0.9624$. The initial
linear power spectrum was obtained from \textsc{Class} and the initial redshift is
$z_\mathrm{ini}=100$. 

\begin{figure}[tbp]
\begin{center}
\includegraphics{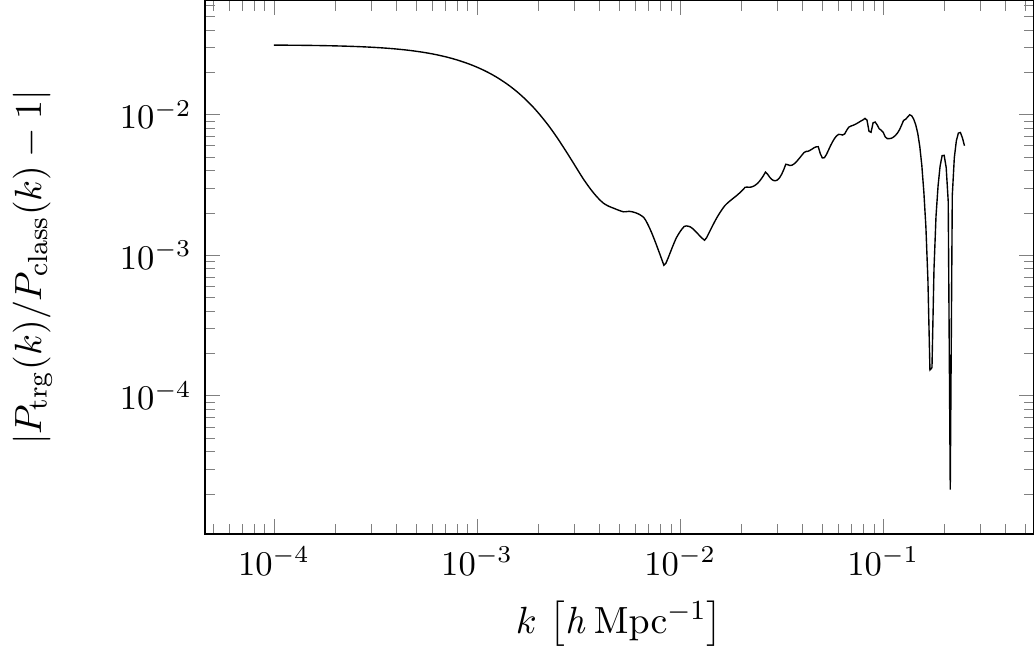}
\end{center}
\caption{The nonlinear power spectrum computed by \textsc{trgfast} at $z=0$ compared
to the one computed by \textsc{Class}. Both agree up to 1\%. }
\label{fig:classcomp}
\end{figure}

Since \textsc{Class} (we use version 2.0) implemented the same algorithm
similarly, we expect to get identical results, and we can in fact
observe that the results are reproduced almost exactly (less than 1\%
difference), as can be seen in \cref{fig:classcomp}. Thus, the entire
analysis in \cite{Audren2011} applies to our code as well.

However, we would expect the difference to vanish completely in the linear
regime, which is not the case here: Even at $k=0.01h/\mathrm{Mpc}$ the
relative difference is around 0.1\%. This appears to be an issue from the
\textsc{Class} implementation, since their own nonlinear power spectrum does
not perfectly match their linear power spectrum at low wave numbers.

Comparing the runtime of \textsc{trgfast} and \textsc{Class} shows an
improvement of a factor of 50 on a common dual-core desktop machine. While
the \textsc{Class} implementation parallelizes on up to 14 cores,
the runtime of \textsc{trgfast} has been shown to be inversely proportional
to the number of cores $n$ until at least $n=24$ and is expected to be able to
use even more cores than that efficiently. 

Another implementation of TRG can be found in the project Copter (we use
version 0.8.7). 
Copter is a C++ library developed by \cite{Carlson2009} where a number of
algorithms from different kinds of perturbation theory are implemented,
including SPT, RPT, LPT, and TRG, the latter being referred to as ``FWT''
(as in ``Flowing With Time'', the title of \cite{Pietroni2008}). The TRG
implementation is very straight forward, resulting in run times of the order
of 30 minutes on a typical dual-core machine despite running in parallel.

However, it needs to be noted that Copter imposes an additional symmetry,
resulting in only 12 instead of 14 independent equations. This is not
mentioned in the accompanying paper, but only in the source code  as a
comment in the file {\tt FlowingWithTime.cpp}:

\begin{quote}
It's not obvious from the definition that $I_{acd,bef}(k)$ is symmetric in
its last two indices $ef$.  This result follows from the fact that it is
initially symmetric ($I_{acd,bef} = 0$ at $\eta = 0$) and the equations of
motion preserve this symmetry.
\end{quote}

This statement appears to be incorrect, considering that the equations of motion
only preserve this symmetry if $A_{acd,bef}$ preserves it, which is only the
case if $P_{11} = P_{12} = P_{22}$. Since the power spectra are only
approximately equal in the linear regime, the symmetry is broken at later
times. Indeed, the resulting nonlinear corrections differ if this symmetry
is assumed. In order to still be able to compare the results, we created
an option to enforce this symmetry in our code.

\begin{figure}[tbp]
\begin{center}
\includegraphics{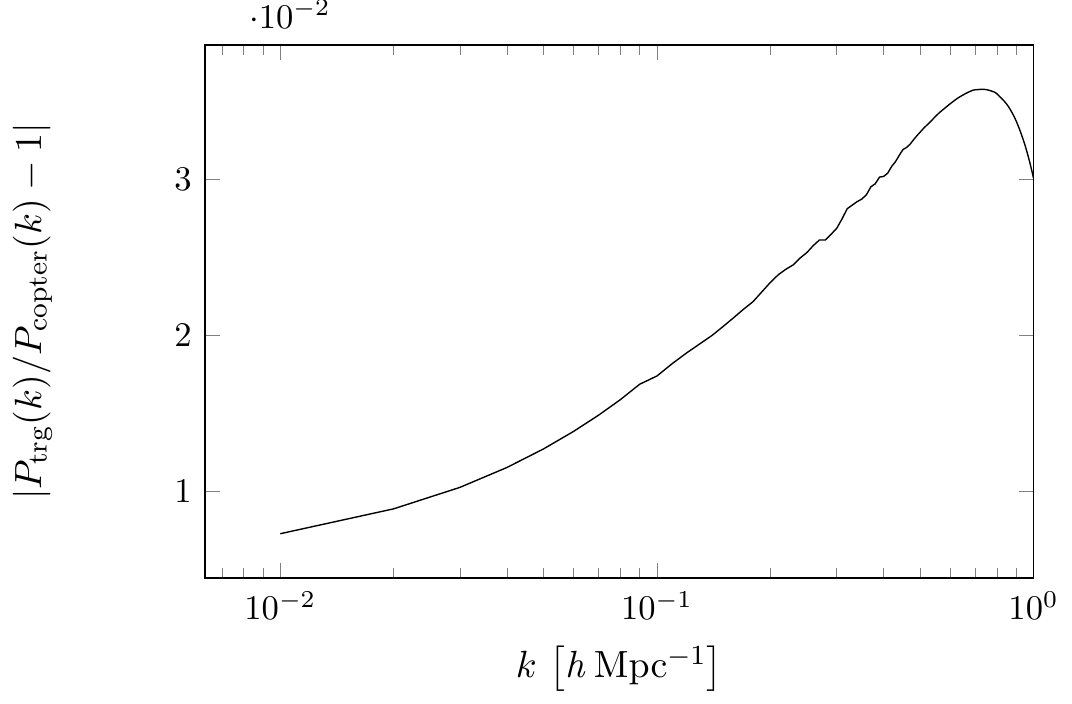}
\end{center}
\caption{The nonlinear power spectrum computed by \textsc{trgfast} at $z=0$ compared
to the one computed by Copter. The power spectra agree up to 3.5\%. }
\label{fig:coptercomp}
\end{figure}

\Cref{fig:coptercomp} shows that the result from \textsc{trgfast} agrees with the
result from Copter at the 3.5\% level. This discrepancy may stem from the
difference in power spectrum extrapolation, cutoff method or numerical
noise in either Copter's code or our code.

\begin{figure}[tbp]
\begin{center}
\includegraphics{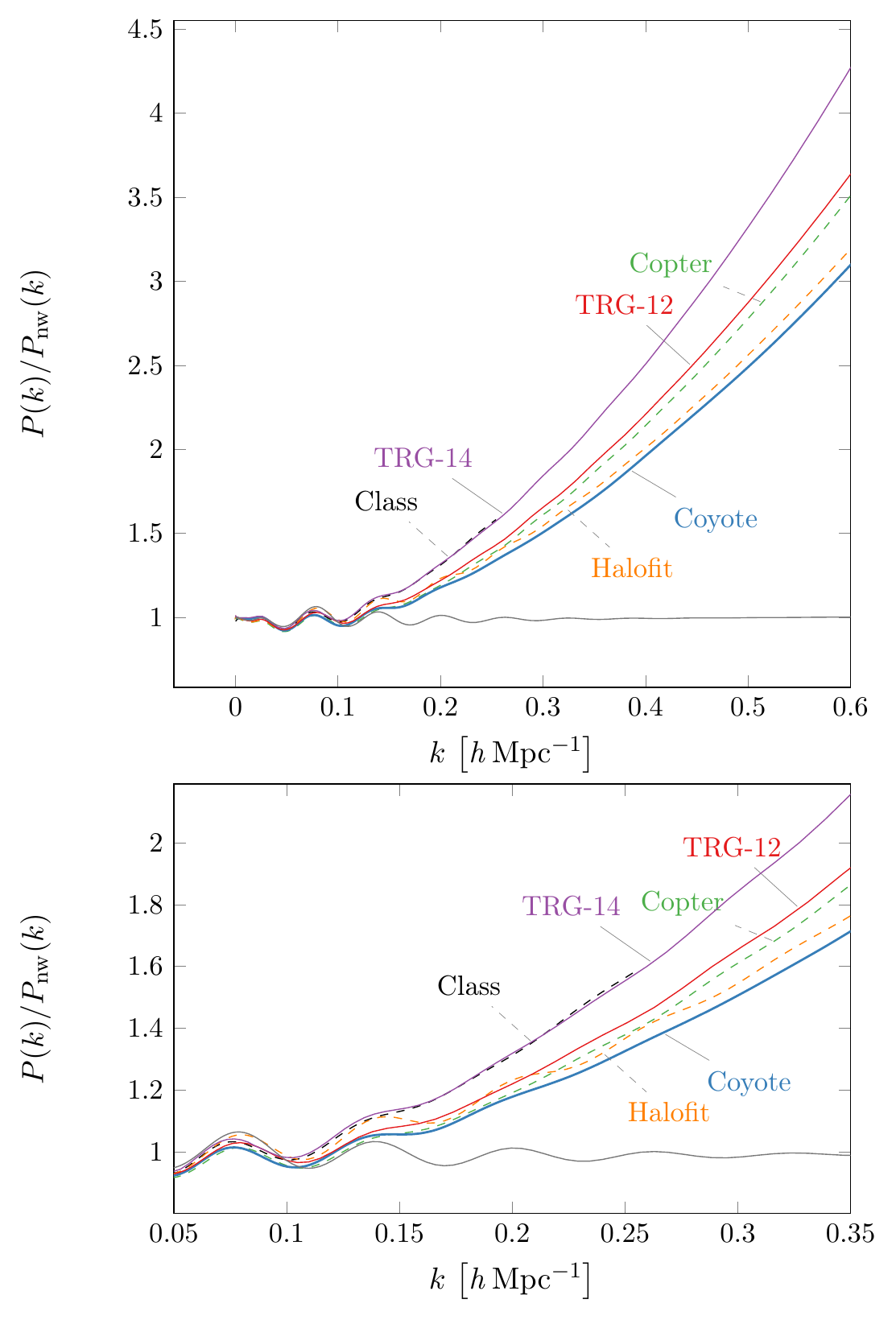}
\end{center}
\caption{A comparison of the power spectrum at $z=0$ obtained from different
sources divided by a linear no-wiggle power spectrum from \cite{Eisenstein1999}.
The sources are Coyote (blue, thick), \textsc{trgfast} with only 12 independent equations
(TRG-12, red), Copter (green, dashed), \textsc{trgfast} with all 14  equations
(TRG-14, purple),
\textsc{Class} (black, dashed), Halofit (orange, dashed), and the linear power
spectrum (gray). The bottom panel is simply a zoomed-in view on the BAO
scale.}
\label{fig:nw-comp}
\end{figure}

An overview of the nonlinear power spectrum at $z=0$ obtained from different
sources can be seen in \cref{fig:nw-comp}.  We notice three pairs of curves:
Unsurprisingly, the Halofit corrections match $N$-body simulations very well
by design even in highly nonlinear scales. The other two pairs correspond to
the power spectrum computed by using the TRG algorithm with and without
enforcing the extra symmetry. Our code can match either curve well while
surpassing the maximum wave number of \textsc{Class} by more than a factor
of 2. For completeness we also included the output from Coyote
\cite{Heitmann2009,Heitmann2010,Lawrence2009,Heitmann2013}, which
interpolates several $N$-body simulations for a \lcdm\ Universe.

\subsection{\texorpdfstring{$N$}{N}-body simulations}

Short of actual data, $N$-body simulations are the best source of the
nonlinear power spectrum in different cosmological models. We will use them
to verify the correctness of our implementation and demonstrate the general
capabilities of the TRG technique.

\subsubsection{\texorpdfstring{\lcdm}{LCDM}}

\begin{figure}[tbp]
\begin{center}
\includegraphics{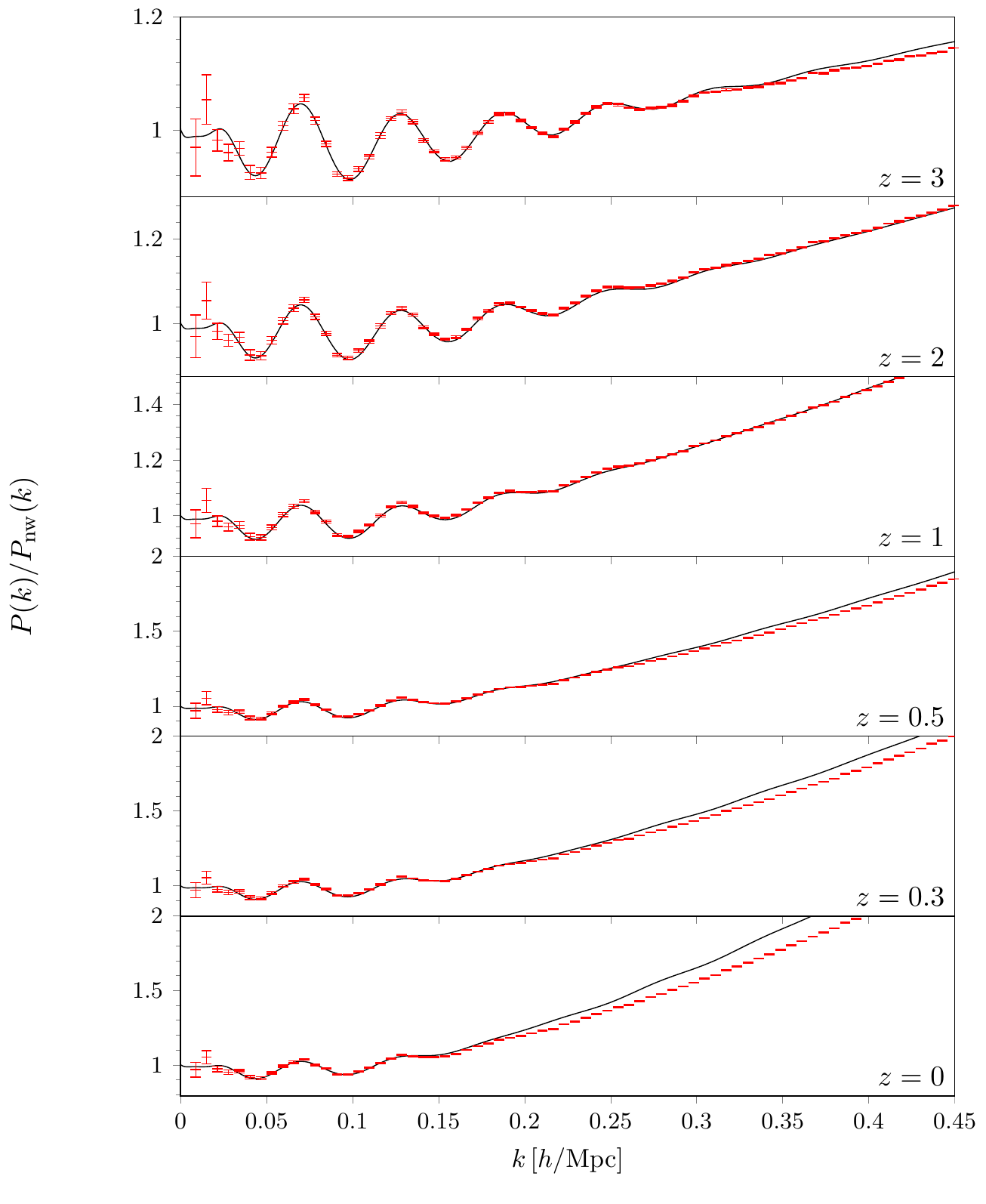}
\end{center}
\caption{Comparison of the \textsc{trgfast} result with \lcdm\ $N$-body data
\cite{Sato2011} for different redshifts. Each power spectrum has been
divided by a no-wiggle spectrum from \cite{Eisenstein1999}. Agreement is
good for redshift $z=1$ and higher, then some discrepancy is noticeable
just as in \cite[fig. 7]{Audren2011}. Note the difference in the scale of
the $y$ direction for some panels!} \label{fig:sato-nbody}
\end{figure}

\begin{figure}[tbp]
\begin{center}
\includegraphics{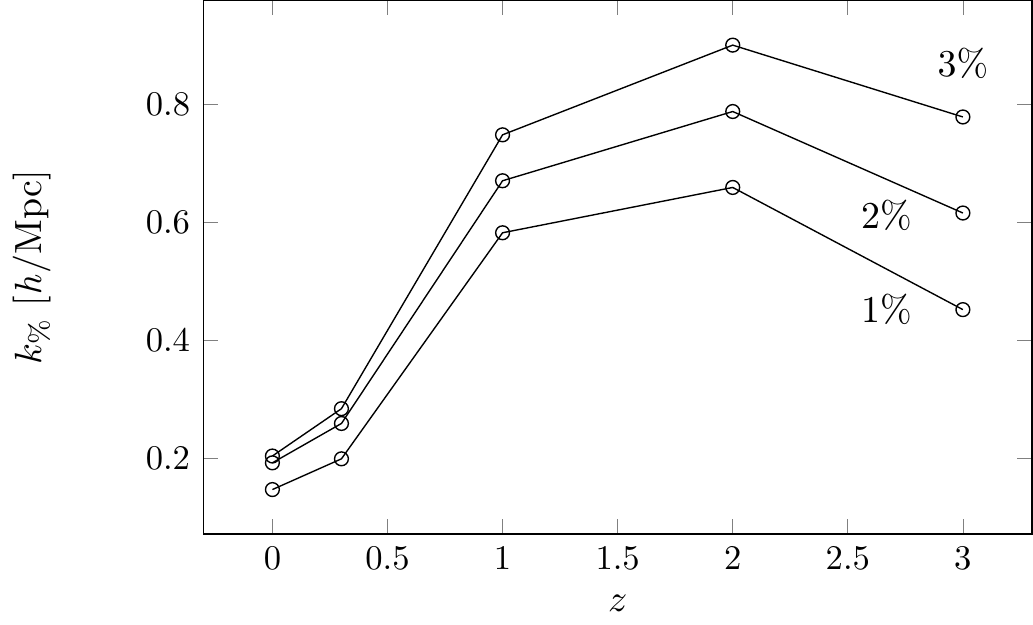}
\end{center}
\caption{These curves indicate at which wave number $k_\%$ the power spectra from
\textsc{trgfast} and \lcdm\ $N$-body simulations \cite{Sato2011} start to deviate by
more than 1\%, 2\%, and 3\% for each redshift. We left out the case for
$z=0.5$ because the power spectra coincidentally agree too well to have a
meaningful value for $k_\%$ here.
\label{fig:sato-nbody-diff}}
\end{figure}

The first choice for a comparison with $N$-body data is the classic flat
\lcdm\ model. We use the data that has been computed by \cite{Sato2011} for
the redshifts $z=3,2,1,0.5,0.3,0$ based on the Wilkinson Anisotropy Probe 7
yr (WMAP7) parameters\footnote{ $\Omega_m=0.265$, $\Omega_b=0.0448$,
$h=0.71$, $\sigma_8=0.80$, $n_s=0.963$} \cite{Komatsu2011}. They ran
\textsc{Gadget2} \cite{Springel2005} with $1024^3$ particles in boxes with
sides $1 h^{-1} \mathrm{Gpc}$. To estimate the uncertainty of the power
spectrum, 30 realizations of the initial conditions based on a linear power
spectrum from \textsc{CAMB} \cite{Lewis1999} were made.
\Cref{fig:sato-nbody} shows their result together with the \textsc{trgfast} power
spectrum for the redshifts mentioned above, both divided by a no-wiggle
power spectrum \cite{Eisenstein1999}. In \cref{fig:sato-nbody-diff}, we
plotted $k_\%$ as a function of the redshift, where $k_\%(z)$ is defined as
the wavenumber where the two power spectra at that redshift first diverge
by more than a given percentage. In the case of $z=0.5$, the power spectra
match so well that $k_\%$ would be unreasonably large, so we omitted that
case.  The TRG result matches the $N$-body data well until the redshift
becomes smaller than unity. This is consistent with \cite[fig.
7]{Audren2011}.  After all, \textsc{trgfast} computes the same quantity as
\textsc{Class} with great precision, as has been shown in
the previous section.

It is somewhat surprising that all three curves in
\cref{fig:sato-nbody-diff} decrease at $z=3$, since we would expect the
power spectra to agree better at higher redshifts, meaning that $k_\%$ would
strictly increase. However, this is in line with \cref{fig:nbody-qc-growth},
where the growth functions from \textsc{trgfast} and $N$-body are compared and the
largest discrepancy is for large redshifts. Of course, that figure is for
coupled quintessence and not \lcdm\ (the linear growth function from
\cite{Sato2011} is unfortunately not available), but qualitative
differences to the \lcdm\ case are unexpected. Ultimately, this comes down
to the fact that by design the growth functions match exactly at $z=0$, so
that any possible deviations have to be at larger redshifts.

\subsubsection{Coupled Quintessence}
\label{sec:cq}

Since one of the advantages of TRG is the flexibility in regard to
cosmological models, we are interested in its performance for models other
than \lcdm.  For a first nonstandard cosmological model, we choose the
coupled quint\-ess\-ence (CQ) model
\cite{Wetterich1994,Amendola1999,Amendola2000}.  
Here, a scalar field $\phi$ couples to ordinary matter via an extra term in
the Langrangian of the form
\[ 
-m_\psi^2 \exp(- C \phi / M_\mathrm{p}^2) \partial_\mu \psi \partial^\mu \psi\,,
\]
where $\psi$ is the ordinary matter field and $C$ the coupling constant. The
Planck mass is $M_\mathrm{p}^2 \equiv 8\pi G$.

Let us consider the energy-momentum tensor $T_{\mu\nu(\phi)}$ for the field
and $T_{\mu\nu(\mathrm{m})}$ for matter. Their sum needs to be locally
conserved, such that we could have
\[ 
\nabla_\mu T^\mu_{\nu(\phi)} = - C T_\mathrm{m} \nabla_\nu \phi,\qquad \nabla_\mu
T^\mu_{\nu(\mathrm{m})} = + C T_\mathrm{m} \nabla_\nu \phi\,.
\]
Other, more complicated  couplings are also possible, but this is the
simplest one and the one we will be considering here. 
For convenience, the coupling constant  is redefined as
\[
\beta  \equiv \sqrt\frac32 M_\mathrm{p}C \,.
\]
As mentioned above, the potential is assumed to read
\[
V(\phi) = V_0 e^{-\sqrt{2/3}  \alpha \phi/M_\mathrm{p}} \,
\]
with the potential parameter $\alpha$.
The evolution of the background functions are derived by
\cite{Amendola2004}.  Using the notation from \cref{sec:trg}, we can
identify 
\[
\mathcal A = -2\beta\sqrt{\Omega_\mathrm{kin}}\,,\quad
\mathcal B = \frac43 \beta^2\,,
\]
such that the background functions  in \cref{eq:bgOmega} take the form 
\begin{align}
\Omega_{21}(a) &= -\frac32 \Omega_\mathrm{m}(a) \left( 1+\frac43\beta^2 \right)\\
\Omega_{22}(a) &= 3 + \frac{\partial \log H(a)}{\partial \log a} -
2\beta\sqrt{\Omega_\mathrm{kin}(a)}\,.
\end{align}
Note that technically we should also include the term \cite{Pietroni2008}
\[ \mathcal B(\vec k,\tau) = \Omega_\mathrm{DE}
\delta^\mathrm{lin}_i/(\Omega_m \delta^\mathrm{lin}_\mathrm{m})\,, \] 
however, we will neglect this term since the quintessence perturbations are
much smaller than the matter perturbations and they are not included in
the $N$-body simulation either.

Several large $N$-body simulations have been performed using the CQ model with
different coupling strengths and different potentials as part of a project
called ``CoDECS''\footnote{Simulation data
publicly available at \url{http://www.marcobaldi.it}.}
\cite{Baldi2010,Baldi2011}. It uses a modified version of \textsc{Gadget-2}
\cite{Springel2005} and features a box size of $L=80h^{-1}\mathrm{Mpc}$ and $2\times
512^3$ particles.  The data set we will be using from CoDECS has the label
``EXP003-L''. Here, the coupling constant is $\beta=0.15$ and the potential
parameter is $\alpha = 0.1$. 

\begin{figure}[tbp]
\begin{center}
\includegraphics{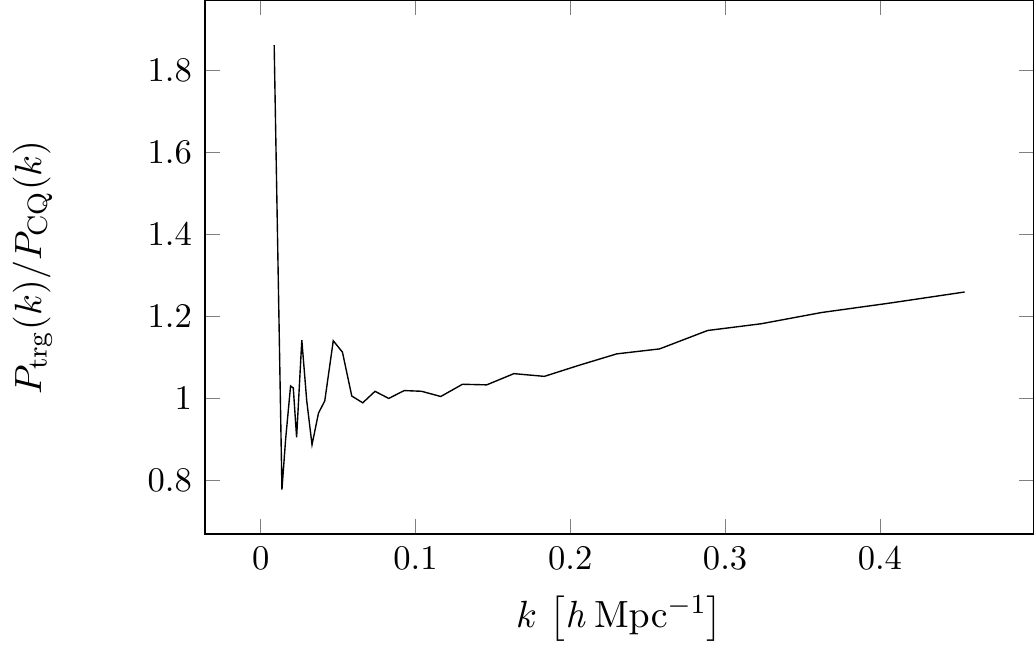}
\end{center}
\caption{The relative difference of the nonlinear power spectrum obtained
from \textsc{trgfast} and from CoDECS at $z=0$ in the coupled quintessence
case. The difference is comparable to the \lcdm\ case.}
\label{fig:nbody-qc}
\end{figure}

We ran the code with the initial linear power spectrum from CoDECS and the
background functions from above. We also use the additional index symmetry
discussed above.  The results are displayed in \cref{fig:nbody-qc}, where
the ratio of the nonlinear power spectrum from \textsc{trgfast} and the
$N$-body simulation has been plotted, both at redshift $z=0$.
Unfortunately, the CoDECS data do not include error bars on the power
spectrum. However, $N$-body simulations naturally cannot constrain the power
spectrum very well on large scales due to their finite volume. This is why
we see large fluctuations for low $k$ in \cref{fig:nbody-qc}. On larger
scales, we have a discrepancy of up to 20\%.

\begin{figure}[tbp]
\begin{center}
\includegraphics{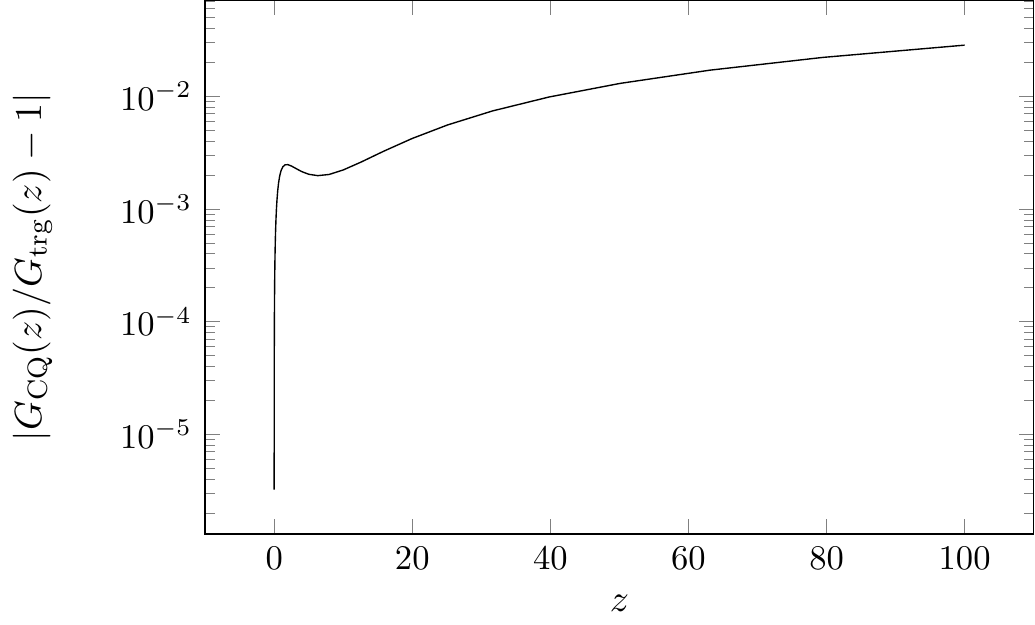}
\end{center}
\caption{The relative difference of the growth function obtained from
\textsc{trgfast} and from CoDECS. The discrepancy is less than 2\% for all relevant
redshifts.}
\label{fig:nbody-qc-growth}
\end{figure}

\Cref{fig:nbody-qc-growth} shows the comparison of the growth functions.
They agree at the 1\% level for all relevant redshifts, which gives reason
to believe that the background evolution was implemented correctly and that
the 20\% discrepancy in the power spectrum has to come from the nonlinear
corrections. The difference is comparable to the one for the \lcdm\ case.

\section{Conclusion}
\label{sec:conclusion}

The TRG framework is one flavor of cosmological perturbation theory that is
applicable to a wider range of models compared to previous perturbation
theories. We showed how we can manipulate the integrand in the integral that
represents the bottleneck of the computation to reduce the 2D integral into
a series of 1D integrals, yielding a speedup of a factor of 50. The trick
consists of expanding the integrand and identifying terms of the shape
$k^mP(k)$, which we call the moments. After semi-analytically integrating
the moments, their antiderivative can be used to perform the first
integration.  The resulting expressions are quite unwieldy and their
representation in the code has been generated automatically by using Wolfram
Mathematica. We ensured the correctness of the code by comparing the output
with the one obtained from \textsc{Class} and Copter as well as $N$-body
simulations in the case of \lcdm\ and coupled quintessence. The maximum
wavenumber at which nonlinear corrections can be produced tends to be higher
than that of the other two projects, but the results are not reliable beyond
the BAO scale.  The code has been released in the public domain and is ready
to be used in other projects.

\section{Acknowledgments}

A.V. and L.A. acknowledge the support from DFG through the project TRR33
``The Dark Universe''.  R.C. acknowledges partial support from the European
Union FP7 ITN INVISIBLES (Marie Curie Actions, PITN-GA-2011-289442) and a
start-up grant funded by the University of G\"ottingen.'' We also thank
Massimo Pietroni for useful discussions.

\bibliographystyle{mine}
\bibliography{bibexport}

\begin{thebibliography}{46}
\newcommand{\enquote}[1]{``#1''}
\providecommand{\natexlab}[1]{#1}
\providecommand{\url}[1]{\texttt{#1}}
\providecommand{\urlprefix}{URL }
\providecommand{\eprint}[2][]{\url{#2}}

\bibitem[{Amara and R\'{e}fr\'{e}gier(2007)}]{Amara2007}
Amara, A. and R\'{e}fr\'{e}gier, A.
\newblock \enquote{{Optimal Surveys for Weak-Lensing Tomography}.}
\newblock \emph{Monthly Notices of the Royal Astronomical Society},
  \textbf{381} (2007)(3), 1018.

\bibitem[{Amendola(1999)}]{Amendola1999}
Amendola, L.
\newblock \enquote{{Coupled Quintessence}.}
\newblock \emph{Physical Review D}, \textbf{62} (1999)(4), 43511.

\bibitem[{Amendola(2000)}]{Amendola2000}
Amendola, L.
\newblock \enquote{{Perturbations in a coupled scalar field cosmology}.}
\newblock \emph{Monthly Notices of the Royal Astronomical Society},
  \textbf{312} (2000)(3), 521.

\bibitem[{Amendola(2004)}]{Amendola2004}
Amendola, L.
\newblock \enquote{{Linear and Nonlinear Perturbations in Dark Energy Models}.}
\newblock \emph{Physical Review D}, \textbf{69} (2004)(10), 103524.

\bibitem[{Amendola et~al.(2013)Amendola, Appleby et~al.}]{Amendola2013b}
Amendola, L., Appleby, S., et~al.
\newblock \enquote{{Cosmology and Fundamental Physics with the Euclid
  Satellite}.}
\newblock \emph{Living Reviews in Relativity}, \textbf{16} (2013).

\bibitem[{Anselmi et~al.(2011)Anselmi, Ballesteros, and Pietroni}]{Anselmi2011}
Anselmi, S., Ballesteros, G., and Pietroni, M.
\newblock \enquote{{Non-linear dark energy clustering}.}
\newblock \emph{Journal of Cosmology and Astroparticle Physics}, \textbf{2011}
  (2011)(11), 014.

\bibitem[{Audren and Lesgourgues(2011)}]{Audren2011}
Audren, B. and Lesgourgues, J.
\newblock \enquote{{Non-linear matter power spectrum from Time Renormalisation
  Group: efficient computation and comparison with one-loop}.}
\newblock \emph{Journal of Cosmology and Astroparticle Physics}, \textbf{2011}
  (2011)(10), 037.

\bibitem[{Baldi(2011)}]{Baldi2011}
Baldi, M.
\newblock \enquote{{Time-dependent couplings in the dark sector: from
  background evolution to non-linear structure formation}.}
\newblock \emph{Monthly Notices of the Royal Astronomical Society},
  \textbf{411} (2011)(2), 1077.

\bibitem[{Baldi et~al.(2010)Baldi, Pettorino et~al.}]{Baldi2010}
Baldi, M., Pettorino, V., et~al.
\newblock \enquote{{Hydrodynamical N -body simulations of coupled dark energy
  cosmologies}.}
\newblock \emph{Monthly Notices of the Royal Astronomical Society},
  \textbf{403} (2010)(4), 1684.

\bibitem[{Bartolo et~al.(2010)Bartolo, Almeida et~al.}]{Bartolo:2009rb}
Bartolo, N., Almeida, J.~B., et~al.
\newblock \enquote{{Signatures of Primordial non-Gaussianities in the Matter
  Power-Spectrum and Bispectrum: the Time-RG Approach}.}
\newblock \emph{JCAP}, \textbf{1003} (2010), 011.
\newblock \eprint{arXiv:0912.4276}.

\bibitem[{Baumann et~al.(2012)Baumann, Nicolis et~al.}]{Baumann:2010tm}
Baumann, D., Nicolis, A., et~al.
\newblock \enquote{{Cosmological Non-Linearities as an Effective Fluid}.}
\newblock \emph{JCAP}, \textbf{1207} (2012), 051.

\bibitem[{Bernardeau and Valageas(2008)}]{Bernardeau:2008ss}
Bernardeau, F. and Valageas, P.
\newblock \enquote{{Propagators in Lagrangian space}.}
\newblock \emph{Phys.Rev.}, \textbf{D78} (2008), 083503.

\bibitem[{Bernardeau et~al.(2002)Bernardeau, Colombi et~al.}]{Bernardeau2002}
Bernardeau, F., Colombi, S., et~al.
\newblock \enquote{{Large-scale structure of the Universe and cosmological
  perturbation theory}.}
\newblock \emph{Physics Reports}, \textbf{367} (2002)(1-3), 1.

\bibitem[{Bernardeau et~al.(2008)Bernardeau, Crocce, and
  Scoccimarro}]{Bernardeau:2008fa}
Bernardeau, F., Crocce, M., and Scoccimarro, R.
\newblock \enquote{{Multi-Point Propagators in Cosmological Gravitational
  Instability}.}
\newblock \emph{Phys.Rev.}, \textbf{D78} (2008), 103521.

\bibitem[{Bernardeau et~al.(2012{\natexlab{a}})Bernardeau, Crocce, and
  Scoccimarro}]{Bernardeau:2011dp}
Bernardeau, F., Crocce, M., and Scoccimarro, R.
\newblock \enquote{{Constructing Regularized Cosmic Propagators}.}
\newblock \emph{Phys.Rev.}, \textbf{D85} (2012{\natexlab{a}}), 123519.

\bibitem[{Bernardeau et~al.(2012{\natexlab{b}})Bernardeau, Van~de Rijt, and
  Vernizzi}]{Bernardeau:2011vy}
Bernardeau, F., Van~de Rijt, N., and Vernizzi, F.
\newblock \enquote{{Resummed propagators in multi-component cosmic fluids with
  the eikonal approximation}.}
\newblock \emph{Phys.Rev.}, \textbf{D85} (2012{\natexlab{b}}), 063509.

\bibitem[{Carlson et~al.(2009)Carlson, White, and Padmanabhan}]{Carlson2009}
Carlson, J., White, M., and Padmanabhan, N.
\newblock \enquote{{Critical look at cosmological perturbation theory
  techniques}.}
\newblock \emph{Physical Review D}, \textbf{80} (2009)(4), 043531.

\bibitem[{Carrasco et~al.(2012)Carrasco, Hertzberg, and
  Senatore}]{Carrasco:2012cv}
Carrasco, J. J.~M., Hertzberg, M.~P., and Senatore, L.
\newblock \enquote{{The Effective Field Theory of Cosmological Large Scale
  Structures}.}
\newblock \emph{JHEP}, \textbf{1209} (2012), 082.

\bibitem[{Crocce and Scoccimarro(2006{\natexlab{a}})}]{Crocce:2005xy}
Crocce, M. and Scoccimarro, R.
\newblock \enquote{{Renormalized cosmological perturbation theory}.}
\newblock \emph{Phys.Rev.}, \textbf{D73} (2006{\natexlab{a}}), 063519.

\bibitem[{Crocce and Scoccimarro(2006{\natexlab{b}})}]{Crocce:2005xz}
Crocce, M. and Scoccimarro, R.
\newblock \enquote{{Memory of initial conditions in gravitational clustering}.}
\newblock \emph{Phys.Rev.}, \textbf{D73} (2006{\natexlab{b}}), 063520.

\bibitem[{Crocce and Scoccimarro(2008)}]{Crocce:2007dt}
Crocce, M. and Scoccimarro, R.
\newblock \enquote{{Nonlinear Evolution of Baryon Acoustic Oscillations}.}
\newblock \emph{Phys.Rev.}, \textbf{D77} (2008), 023533.

\bibitem[{Crocce et~al.(2012)Crocce, Scoccimarro, and Bernardeau}]{Crocce2012}
Crocce, M., Scoccimarro, R., and Bernardeau, F.
\newblock \enquote{{MPTbreeze: A fast renormalized perturbative scheme}.}
\newblock \textbf{10} (2012)(July), 16.

\bibitem[{Eisenstein and Hu(1999)}]{Eisenstein1999}
Eisenstein, D.~J. and Hu, W.
\newblock \enquote{{Power Spectra for Cold Dark Matter and Its Variants}.}
\newblock \emph{The Astrophysical Journal}, \textbf{511} (1999)(1), 5.

\bibitem[{Heitmann et~al.(2009)Heitmann, Higdon et~al.}]{Heitmann2009}
Heitmann, K., Higdon, D., et~al.
\newblock \enquote{{The Coyote Universe. II. Cosmological Models and Precision
  Emulation of the Nonlinear Matter Power Spectrum}.}
\newblock \emph{The Astrophysical Journal}, \textbf{705} (2009)(1), 156.

\bibitem[{Heitmann et~al.(2010)Heitmann, White et~al.}]{Heitmann2010}
Heitmann, K., White, M., et~al.
\newblock \enquote{{The Coyote Universe. I. Precision Determination of the
  Nonlinear Power Spectrum}.}
\newblock \emph{The Astrophysical Journal}, \textbf{715} (2010)(1), 104.

\bibitem[{Heitmann et~al.(2013)Heitmann, Lawrence et~al.}]{Heitmann2013}
Heitmann, K., Lawrence, E., et~al.
\newblock \enquote{{The Coyote Universe Extended: Precision Emulation of the
  Matter Power Spectrum}.}
\newblock \emph{arXiv},  (2013), 17.

\bibitem[{Huterer(2002)}]{Huterer2002}
Huterer, D.
\newblock \enquote{{Weak Lensing and Dark Energy}.}
\newblock \emph{Physical Review D}, \textbf{65} (2002)(6), 063001.

\bibitem[{Komatsu et~al.(2010)Komatsu, Smith et~al.}]{Komatsu2011}
Komatsu, E., Smith, K.~M., et~al.
\newblock \enquote{{Seven-Year Wilkinson Microwave Anisotropy Probe ({WMAP})
  Observations: Cosmological Interpretation}.}
\newblock \emph{The Astrophysical Journal Supplement Series}, \textbf{192}
  (2010)(2), 57.

\bibitem[{Laureijs et~al.(2011)Laureijs, Amiaux et~al.}]{Laureijs2011}
Laureijs, R., Amiaux, J., et~al.
\newblock \enquote{{Euclid Definition Study Report}.}
\newblock Technical report, Euclid collaboration (2011).

\bibitem[{Lawrence et~al.(2010)Lawrence, Heitmann et~al.}]{Lawrence2009}
Lawrence, E., Heitmann, K., et~al.
\newblock \enquote{{The Coyote Universe. III. Simulation Suite and Precision
  Emulator for the Nonlinear Matter Power Spectrum}.}
\newblock \emph{The Astrophysical Journal}, \textbf{713} (2010)(2), 1322.

\bibitem[{Lesgourgues and Pastor(2006)}]{Lesgourgues2006}
Lesgourgues, J. and Pastor, S.
\newblock \enquote{{Massive neutrinos and cosmology}.}
\newblock \emph{Physics Reports}, \textbf{429} (2006)(6), 307.

\bibitem[{Lesgourgues et~al.(2009)Lesgourgues, Matarrese
  et~al.}]{Lesgourgues:2009am}
Lesgourgues, J., Matarrese, S., et~al.
\newblock \enquote{{Non-linear Power Spectrum including Massive Neutrinos: the
  Time-RG Flow Approach}.}
\newblock \emph{JCAP}, \textbf{0906} (2009), 017.

\bibitem[{Lewis et~al.(2000)Lewis, Challinor, and Lasenby}]{Lewis1999}
Lewis, A., Challinor, A., and Lasenby, A.
\newblock \enquote{{Efficient Computation of Cosmic Microwave Background
  Anisotropies in Closed Friedmann-Robertson-Walker Models}.}
\newblock \emph{The Astrophysical Journal}, \textbf{538} (2000)(2), 473.

\bibitem[{Manzotti et~al.(2014)Manzotti, Peloso et~al.}]{Manzotti:2014loa}
Manzotti, A., Peloso, M., et~al.
\newblock \enquote{{A coarse grained perturbation theory for the Large Scale
  Structure, with cosmology and time independence in the UV}.}
\newblock \emph{JCAP}, \textbf{1409} (2014)(09), 047.

\bibitem[{Matarrese and Pietroni(2007)}]{Matarrese:2007wc}
Matarrese, S. and Pietroni, M.
\newblock \enquote{{Resumming Cosmic Perturbations}.}
\newblock \emph{JCAP}, \textbf{0706} (2007), 026.

\bibitem[{Matsubara(2008)}]{Matsubara:2007wj}
Matsubara, T.
\newblock \enquote{{Resumming Cosmological Perturbations via the Lagrangian
  Picture: One-loop Results in Real Space and in Redshift Space}.}
\newblock \emph{Phys.Rev.}, \textbf{D77} (2008), 063530.

\bibitem[{Piazza and Vernizzi(2013)}]{Piazza:2013coa}
Piazza, F. and Vernizzi, F.
\newblock \enquote{{Effective Field Theory of Cosmological Perturbations}.}
\newblock \emph{Class.Quant.Grav.}, \textbf{30} (2013), 214007.

\bibitem[{Pietroni(2008)}]{Pietroni2008}
Pietroni, M.
\newblock \enquote{{Flowing with time: a new approach to non-linear
  cosmological perturbations}.}
\newblock \emph{Journal of Cosmology and Astroparticle Physics}, \textbf{2008}
  (2008)(10), 036.

\bibitem[{Pietroni et~al.(2012)Pietroni, Mangano et~al.}]{Pietroni:2011iz}
Pietroni, M., Mangano, G., et~al.
\newblock \enquote{{Coarse-Grained Cosmological Perturbation Theory}.}
\newblock \emph{JCAP}, \textbf{1201} (2012), 019.

\bibitem[{Pueblas and Scoccimarro(2009)}]{Pueblas2009}
Pueblas, S. and Scoccimarro, R.
\newblock \enquote{{Generation of vorticity and velocity dispersion by orbit
  crossing}.}
\newblock \emph{Physical Review D}, \textbf{80} (2009)(4), 043504.

\bibitem[{Saracco et~al.(2010)Saracco, Pietroni et~al.}]{Saracco:2009df}
Saracco, F., Pietroni, M., et~al.
\newblock \enquote{{Non-linear Matter Spectra in Coupled Quintessence}.}
\newblock \emph{Phys.Rev.}, \textbf{D82} (2010), 023528.

\bibitem[{Sato and Matsubara(2011)}]{Sato2011}
Sato, M. and Matsubara, T.
\newblock \enquote{{Nonlinear biasing and redshift-space distortions in
  Lagrangian resummation theory and N-body simulations}.}
\newblock \emph{Physical Review D}, \textbf{84} (2011)(4), 1.

\bibitem[{Springel(2005)}]{Springel2005}
Springel, V.
\newblock \enquote{{The cosmological simulation code GADGET-2}.}
\newblock \emph{Monthly Notices of the Royal Astronomical Society},
  \textbf{364} (2005)(4), 1105.

\bibitem[{Taruya et~al.(2012)Taruya, Bernardeau et~al.}]{Taruya:2012ut}
Taruya, A., Bernardeau, F., et~al.
\newblock \enquote{{RegPT: Direct and fast calculation of regularized
  cosmological power spectrum at two-loop order}.}
\newblock \emph{Phys.Rev.}, \textbf{D86} (2012), 103528.

\bibitem[{Valageas(2010)}]{Valageas2010}
Valageas, P.
\newblock \enquote{{Impact of shell crossing and scope of perturbative
  approaches, in real and redshift space}.}
\newblock \emph{Astronomy \& Astrophysics}, \textbf{526} (2010), A67.

\bibitem[{Wetterich(1994)}]{Wetterich1994}
Wetterich, C.
\newblock \enquote{{An asymptotically vanishing time-dependent cosmological
  "constant"}.}
\newblock \emph{Astronomy and Astrophysics}, \textbf{301} (1994), 321.

\end{thebibliography}

\end{document}